\documentclass[aip,reprint,twocolumn,notitlepage,secnumarabic,amssymb, nobibnotes, superscriptaddress]{revtex4-1}
\usepackage{graphicx}
\usepackage{wrapfig}
\usepackage{bm}
\usepackage{float}
\usepackage{braket}
\usepackage{upgreek}
\usepackage{physics}
\begin{document}
\preprint{AIP/123-QED}
\title{The role of the quadrupolar interaction in the tunneling dynamics of lanthanide molecular magnets}%

\author{Gheorghe Taran}%
\email[]{gheorghe.taran@kit.edu}
\affiliation{Physikalisches Institute, KIT, Wolfgang-Gaede-Str. 1, Karlsruhe D-76131}
\author{Edgar Bonet}%
\email[]{bonet@grenoble.cnrs.fr}
\affiliation{N\'eel Institute, CNRS, 25 rue des Martyrs,
Grenoble 38042}

\author{Wolfgang Wernsdorfer}%
\email[]{wolfgang.wernsdorfer@kit.edu}
\affiliation{Physikalisches Institute, KIT, Wolfgang-Gaede-Str. 1, Karlsruhe D-76131}
\affiliation{N\'eel Institute, CNRS, 25 rue des Martyrs,
Grenoble 38042}
\affiliation{Institute of Nanotechnology (INT), Karlsruhe Institute of Technology (KIT), Hermann-von-Helmholtz-Platz 1, D-76344 Eggenstein-Leopoldshafen}

\date{\today}%

\begin{abstract}
Quantum tunneling dominates the low temperature magnetization dynamics in molecular magnets and presents features that are strongly system dependent.
The current discussion is focused on the terbium(III) bis(phtalocyanine) ([TbPc$_2$]$^{-1}$) complex, that should serve as a prototypical case for lanthanide molecular magnets.   
We analyze numerically the effect of non-axial interactions on the magnitude of the intrinsic tunnel splitting and show that usual suspects like the transverse ligand field and Zeeman interaction fail to explain the experimentally observed dynamics. 
We then propose through the nuclear quadrupolar interaction a viable mechanism that mixes, otherwise \textit{almost} degenerate hyperfine states.  
\end{abstract}

\maketitle

\section{Introduction} 
Single ion molecular magnets (SIMMs) are the first members in the ever growing family of magnets that employ molecules as their basic unit~\cite{zheng2014}.
Their simple magnetic core motivated a considerable scientific effort to correlate different chemical designs to exhibited magnetic properties~\cite{luzon2012lanthanides, tang2016}, and to induce and control coherent quantum dynamics on a large timescale~\cite{ ardavan2007, zadrozny2015, atzori2016}.
The objective is to reach a level of understanding that will allow to synthesize molecular units suited to be incorporated in functional devices~\cite{ganzhorn2014molecular}.
In the search for the optimal molecular design (to enhance bistability and dynamical properties of SIMMs)
the chemical control over
the symmetry of the complex and the nature of the ligand substituents proved to be essential~\cite{liu2018symmetry,mcadams2017molecular}. 
The progress made in the last decades being relevant both to the fundamental research in the field of mesoscopic quantum physics~\cite{barbara2014quantum},
and to the trend and outlook of the current technology~\cite{stamp2009, troiani2011molecular, ghirri2017molecular}.
\begin{figure}
    \centering
    \includegraphics[width=0.48\textwidth]{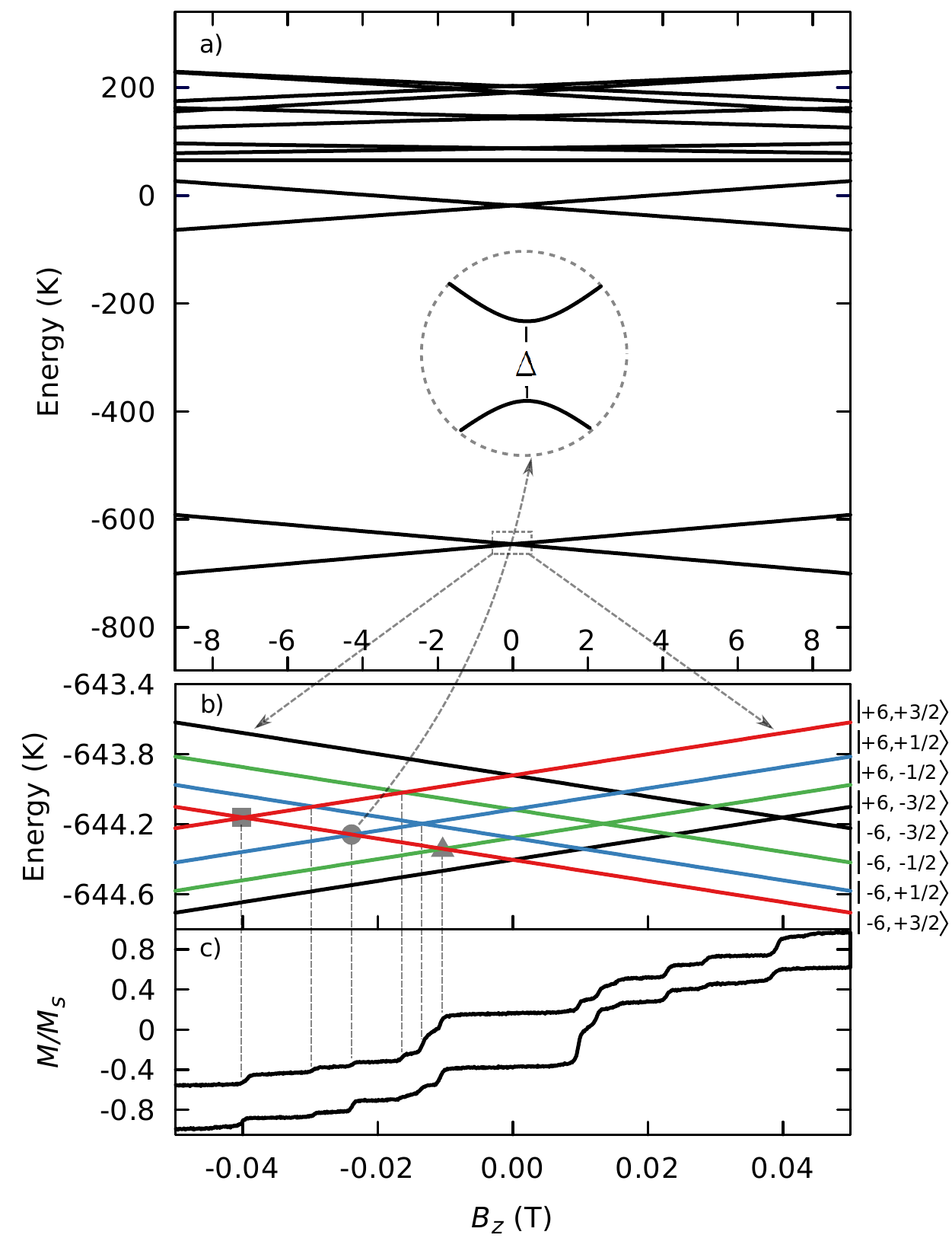}
    \caption{
    \label{fig:zd}
    \text{a)} Zeeman diagram that corresponds to the Hamiltonian given by Eq.~(\ref{eq:H}). The inset shows the tunnel splitting between two mixed hyperfine states.
    \text{b)} Hyperfine structure of the lowest doublet, $m_J = \pm 6$ in the field region where tunneling transitions take place.
    \text{c)} Magnetic hysteresis loop of TbPc$_2$ complex in a diluted single crystal. The position of the relaxation steps are fitted to the corresponding level crossings in the Zeeman diagram by using $A_{\text{hyp}} = 26.7$ mK and $D_{\text{quad}}=17$~mK.
    }
    \vspace{-15pt}
\end{figure}

A central property of molecular magnets is their magnetic bistability, that is, the existence of an energy barrier that separates states of different spin orientation.
Thus, they are envisioned as memory units in high density storage devices~\cite{affronte2014potentialities}.
The obvious strategy to reach this highly sought goal is to enhance the uniaxial anisotropy to obtain molecular complexes that exhibit hysteresis at high enough temperatures.
Advancement in this direction has been recently reported as a mononuclear Dy compound was shown to exhibit magnetic hysteresis at temperatures up to 60 K~\cite{guo2017dysprosium,goodwin2017molecular}.
However many challenges still need to be surmounted, one of which lies in the intrinsic quantum nature of molecular magnets itself.
Notably, underbarrier relaxation pathways (\textit{e.g.} pure and phonon assisted quantum tunneling), opened by transverse interactions that break the axial symmetry of the molecule, results in a much lower effective energy barrier.

Quantum tunneling of magnetization  was also instrumental in reading out and manipulating
both the electronic and the nuclear spin of a mononuclear molecular complex~\cite{thiele2014electrically},
to the point of the successful implementation of quantum algorithms~\cite{godfrin2017operating}.
Thus, understanding the mechanisms that operate behind the observed tunneling dynamics is an important prerequisite to design application oriented molecular magnets.

Amongst already numerous example of SIMMs, the TbPc$_2$ molecule can be linked to breakthrough discoveries that greatly helped to advance the agenda of this research field. 
First, through ac-measurements it was noticed that a molecule with a single magnetic center can exhibit a large energy barrier~\cite{ishikawa2003lanthanide}.
Then, micro-SQUID measurements on a diluted TbPc$_2$ molecular crystal experimentally showed resonant relaxation through quantum tunneling between mixed states of both electronic and nuclear origin~\cite{ishikawa2005quantum}.
Thus, it became the system of choice for the first molecular spintronics devices~\cite{komeda2011observation, urdampilleta2011supramolecular,vincent2012electronic} and it helped to construct the case for using molecules as potential qubits.

In this communication we revisit the analysis of the low temperature hysteresis loop (Fig.~\ref{fig:zd}c) characterizing a diluted crystal of [TbPc$_2$]$^{-1}$ SIMMs in an isostructural diamagnetic YPc$_2$ matrix
(Tb to Y ration of 1\%~\cite{konami1989analysis}),
measured with the micro-SQUID setup at subkelvin temperatures~\cite{wernsdorfer2009}.
We show that despite the great progress made in the last decade and numerous studies that looked at the TbPc$_2$ complex~\cite{ishikawa2010functional}, the tunneling dynamics of this system was still poorly understood.
We investigate numerically different interactions that have the potential to promote tunneling transitions and show that the coupling of the electronic shell to the $^{159}$Tb nuclear spin dominates the environmental and ligand field interactions.

\begin{figure}
    \includegraphics[width=0.48\textwidth]{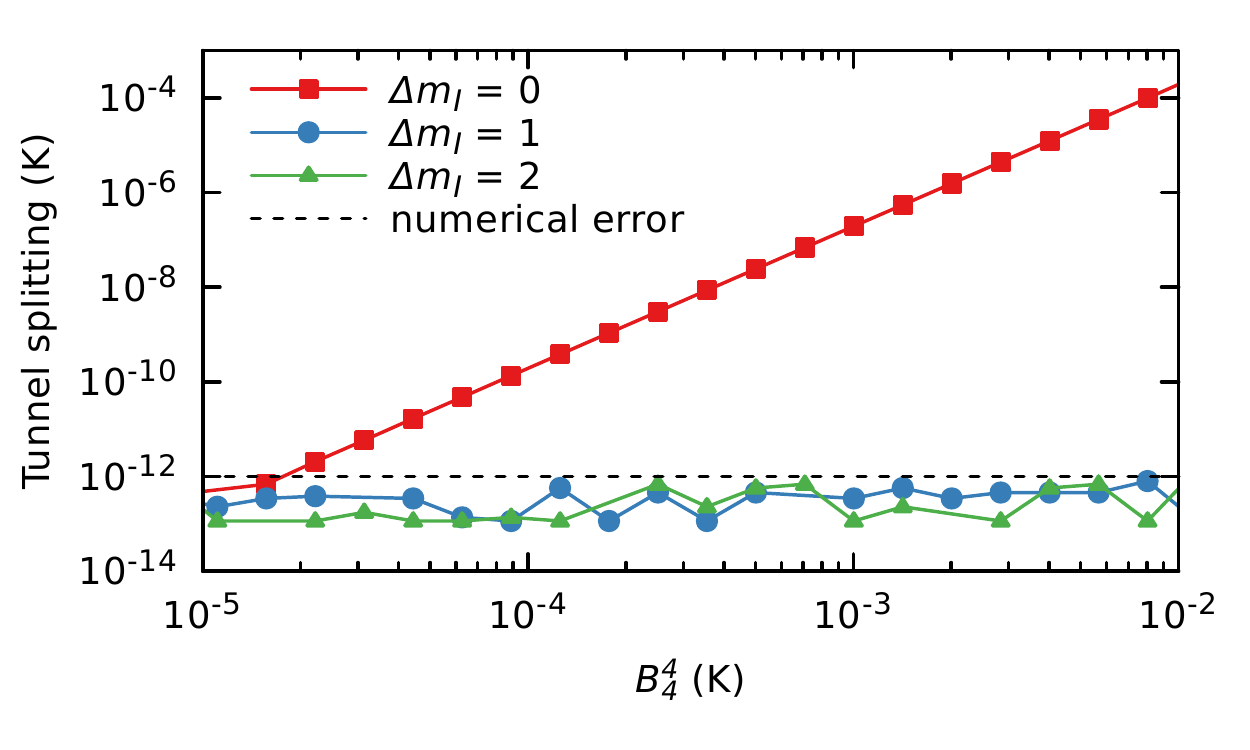}
    \caption{
    \label{fig:b44}
    The dependence of the tunnel splitting of the three types of anticrossings (denoted by the change of the nuclear spin, $\Delta m = 0, 1\text{ and } 2$)  on the amplitude of the $O_4^4$ term in the ligand field interaction (Eq.~\ref{eq:lf}). Only the transitions that conserve the nuclear spin are allowed while all the other level crossings should be degenerate.
    }
\end{figure}

\section{Theory} 
The Tb$^{3+}$ ion, 
found at the core of the molecule, has 
the [Xe]4f$^8$ ground electronic structure with 
a total angular momentum ($J$) taking values between 0 and 6.
The strong spin-orbit interaction leads to 
a separation of about 2900 K between the ground ($J=6$) and first excited multiplet ($J=5$). 
The Tb$^{3+}$ ion can be embedded between two parallel phtalocyanine (Pc) ligand planes as it coordinates with 4 nitrogens from each plane in a square antiprismatic (D$_{\text{4d}}$) geometry.
The crystal field interaction generates a quantization axis oriented perpendicular to the Pc planes and further splits the $2J+1$ states of the ground multiplet.

The ligand field interaction was modeled, with a certain degree of success, by the following Hamiltonian~\cite{ishikawa2005quantum}:
\begin{equation}
    \label{eq:lf}
    \mathcal{H}_{\text{lf}} =  \sum_{n=1}^3 B_{2n}^0 O_{2n}^0 +
    B_4^4  O_4^4
    \end{equation}
where $O_{q}^k$ are the extended Steven's operators~\cite{abragam1970} and $B_{q}^k$ are the ligand field parameters obtained by performing a simultaneous fit of ac-susceptibility and $^1$H-NMR  measurements of the isostructural lanthanide double decker series~\cite{ishikawa2003lanthanide}.
The terms in the sum describe the uniaxial anisotropy while the $O_4^4$ term models a transverse ligand field interaction that arises from a broken D$_{\text{4d}}$ symmetry. 

Another important factor determining the electronic structure of the compound is found in 
the strong hyperfine interaction, $ A_{\text{hyp}}\mathbf{I} \cdot \mathbf{J}$, between the electronic shell and 
the Tb nucleus ($^{159}$Tb isotope with a natural abundance of 100~\%), resulting in 
an effective half integer spin.
Non-spherical charge distribution around the $^{159}$Tb nucleus with a spin angular momentum $I=3/2$ gives 
a non-negligible quadrupolar contribution, $\mathbf{I}\hat{D}_{\text{quad}}\mathbf{I}$. 
In the ideal case of a D$_{\text{4d}}$ symmetry of the electronic shell, only the axial term ($\sim I_z^2$) needs to be preserved.
Thus, the total Hamiltonian that also includes the coupling to a magnetic field ($\mathbf{H}$), is given by the following expression:
\begin{align}
\label{eq:H}
\mathcal{H}_{\text{TbPc$_{2}$}}  
&=  \mathcal{H}_{\text{lf}} + \mu_{\text{B}} \mu_{\text{0}} \mathbf{H}\cdot (g_e \mathbf{J} + g_n \mathbf{I})\nonumber\\ 
&+  A_{\text{hyp}}\mathbf{I} \cdot \mathbf{J} +
\mathbf{I}\hat{D}_{\text{quad}}\mathbf{I}
\end{align}
where the second term is the Zeeman interaction parametrized through 
the  electronic ($g_e = 1.5$) and the nuclear ($g_n = 1.33$) gyromagnetic ratios. 
The $A_{\text{hyp}}$ and axial (dominant) term of $\hat{D}_{\text{quad}}$ ($D_{\text{quad}}$)  are uniquely determined by 
the positions of the relaxation steps (Fig.~\ref{fig:zd}c) that are fitted to the corresponding level crossings in the Zeeman diagram (Fig.~\ref{fig:zd}b), leading to $A_{\text{hyp}} = 26.7$ mK and $D_{\text{quad}}=17$~mK.  

\section{Results and Discussion}
\begin{figure}
    \centering
    \begin{minipage}{0.48\textwidth}
        \includegraphics[width=.99\textwidth]{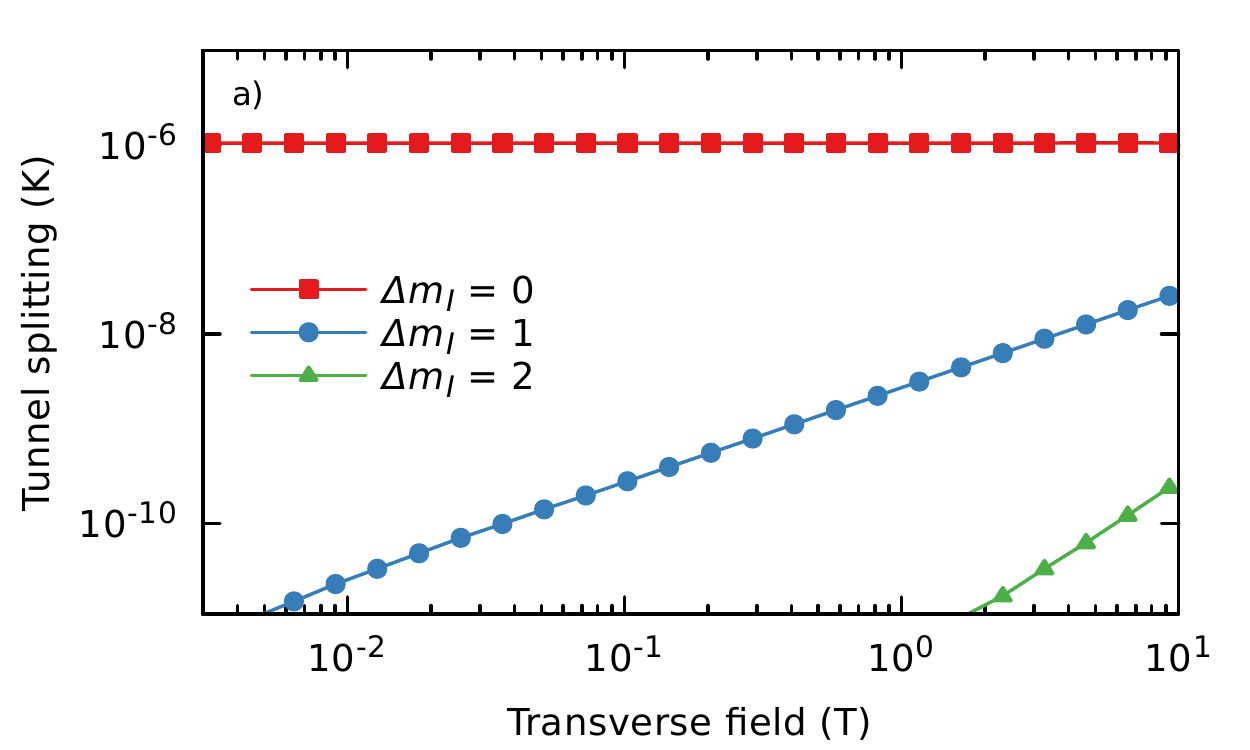}
    \end{minipage}
    \vfill
    \begin{minipage}{0.48\textwidth}
        \includegraphics[width=.99\textwidth]{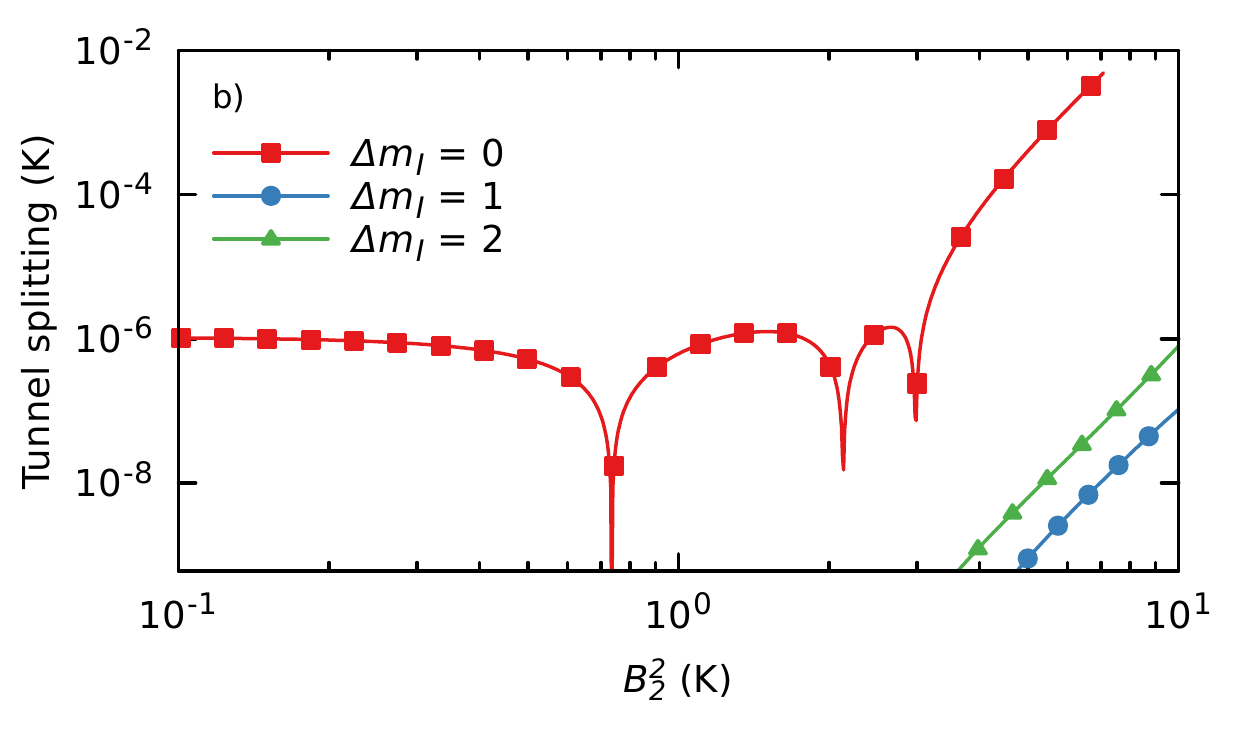}
    \end{minipage}
    \caption{
    \label{fig:ht}
    a) The effect of an applied transverse field on the tunnel splitting of the different hyperfine transitions. The transverse Hamiltonian includes also the fourth order perturbation with $B_4^4 = 6$ mK.
    b) An additional rhombic term ($B_2^2O_2^2$) is added to the situation shown in the above panel, where we consider a constant transverse field, $H_x = 1$ mT. 
    The observed oscillations are the result of the Berry phase interference~\cite{wernsdorfer1999}.
    }
\end{figure}
Figure \ref{fig:zd}c  shows a zoom of the magnetic hysteresis loop, measured by using the microSQUID technique~\cite{wernsdorfer2009}, of a crystal containing TbPc$_2$ SIMMs diluted in a diamagnetic, isostructural matrix formed by YPc$_2$ molecules, with [TbPc$_2$]/[YPc$_2$] ratio of 1~\%. Upon sweeping the magnetic field from $-1$ T up to positive fields as small as 0.05 T, approximately 75~\% of the TbPc$_2$ SIMMs undergo quantum tunneling transitions, resulting in sharp steps in the magnetization curve. The remaining SMM reverse their magnetic moment at larger magnetic fields by a direct relaxation process~\cite{ganzhorn2014molecular}.
Quantum tunnel transitions take place between the mixed states of nuclear and electronic origin, thus both spin projections can change. 
The tunnel splitting ($\Delta$ in the inset of Fig.~\ref{fig:zd}a) quantifies the state mixing at resonance
and is directly connected to the 
magnitude of the tunneling transition rates~\cite{gatteschi2006}.
Thus, the main objective of this paper is to evaluate $\Delta$ using numerical methods and this way to single out the interactions that promote tunneling relaxation between different hyperfine states.

The first observation we make, concerns the absence of a relaxation step at 
the level crossings found at zero external field. 
This is explained through Kramer's theory for half integer spin system that predicts 
degenerate ground states if only the ligand field is taken into account~\cite{wernsdorfer2005quantum}. 
The rest of the transitions can be labeled by the change in the nuclear magnetic moment ($\Delta m_I = 0,1 \text{ and }2$, shown in Fig.~\ref{fig:zd}b as a square, circle, and triangle, respectively). 

Factors that lead to level mixing are the ones that break the axial symmetry of the system. 
For example, $O_4^4$ operator in the ligand field Hamiltonian mixes $\pm 6$ electronic states and transitions between them become allowed. 
By varying the $B^4_4$ 
parameter between $10^{-5}$ K and $10^{-2}$ K, the tunnel splitting corresponding to the crossings that conserves the nuclear spin vary between $10^{-12}$ K and $10^{-4}$ K, while all the other splittings remain negligible (Fig.~\ref{fig:b44}). 
This is not a surprising result as a fourth order perturbation can induce transitions only between states with a total spin that differ by a multiple of four.
If this would be the sole non-axial interaction, only the transitions that conserve the nuclear spin will be observed -- which is not our case.

One factor that is often invoked when explaining why the 
selection rules are not obeyed is the transverse component of the magnetic field that can be of both internal (\textit{e.g.} dipolar field) or external (\textit{e.g.} applied field) origin.
When computing the tunnel splitting as a function of the applied transverse field (Fig.~\ref{fig:ht}a), one observes that a field of at least 1~T is needed in order to have significant splittings pertaining to the crossings that do not conserve the nuclear spin (taking the ones that conserve the nuclear spin as a reference).
This value is of course much larger than the environmental fields of dipolar origin. The dipolar field variance for our sample of 1~\% concentration being around $\sqrt{\left<\Delta H_{\text{dip}}^2\right>} \approx 1$~mT.
The necessity for such large values is easy to understand as the coupling between the magnetic field and the nuclear spin is weak due to the smallness of the nuclear magnetic moment.
\newline
The deviation from the tetragonal symmetry by the inclusion of a second order perturbation ($B_2^2O_2^2$) was already done when the low temperature tunneling dynamics of Mn$_{12}$-ac was analyzed~\cite{del2005magnetic}.
If one were to add also the biaxial term to the ligand field Hamiltonian (Eq.~\ref{eq:lf}), the predicted dynamics of 
the $|\Delta m| = 1\text{ or } 2$ transitions would still remain 
orders of magnitude slower than the relaxation at the crossings that conserve the nuclear spin (Fig.~\ref{fig:ht}b).
  
The reason why the above factors fail to explain the observed transitions lies in
the magnitude of the different interactions described by Eq.~\ref{eq:H}.
The effective total half integer spin comes from two heterogeneous spins that have a strong uniaxial anisotropy and are tightly coupled through the hyperfine interaction. 
The hyperfine interaction does not promote transitions of both the electronic and nuclear spin (through terms like $A_{\text{hyp}}J_+I_-$) because the selection rules cannot be satisfied by both a nuclear and a electronic spin transition.
While the electronic states are easily mixed by
the ligand field, 
the nuclear states, to the first order, remain degenerate.
Thus, in order to explain the observed steps in the magnetization curve, one needs to look at interactions that strongly couple to the nuclear spin.
 
\begin{figure}
    \includegraphics[width=0.48\textwidth]{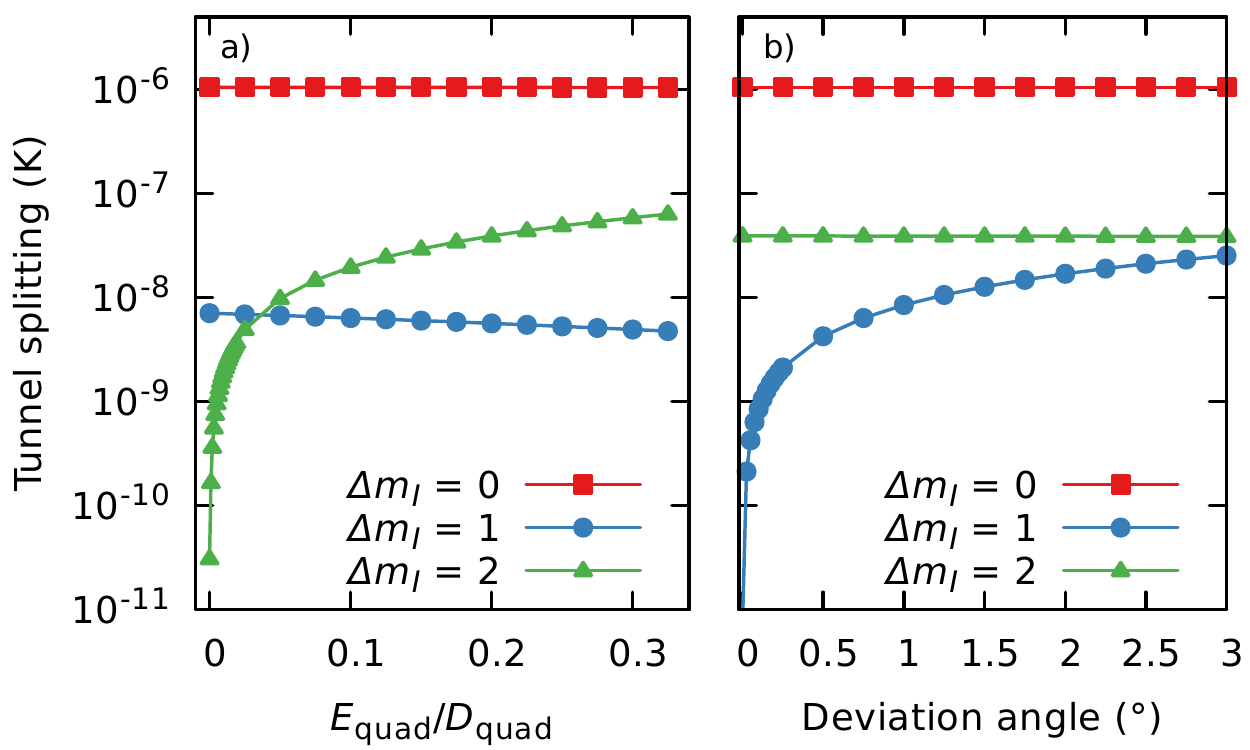}
    \caption{
    \label{fig:quad}
    \textbf{(a)} Variation of the tunnel splitting with the rhombic term ($E_{\text{quad}}(I_x^2-I_y^2)$) in the quadrupolar interaction, computed for a fixed angle misalignment between the ligand field and quadrupolar easy axis, of 1$^\circ$. From other transverse interactions only the $O_4^4$ term was kept, with $B_4^4 = 0.5$ K.
    \textbf{(b)} Variation of the tunnel splitting as a function of the misalignment between the ligand field and quadrupolar easy axis for $E_{\text{quad}} / D_{\text{quad}} = 0.2$, and the same $O_4^4$ term.
	}
\end{figure}

We suggest that a solution can be found in 
the quadrupolar interaction between the nucleus and the electronic shell.
We already mentioned that if we consider a broken square antiprismatic symmetry we can add a biaxial term to the ligand field Hamiltonian. 
This entails us to include the biaxial term to the quadrupolar Hamiltonian as well, thus:  $\mathbf{I}\hat{D}_{\text{quad}}\mathbf{I} = D_{\text{quad}} I_z^2 + E_{\text{quad}}(I_x^2-I_y^2)$. 
With this term, the states of the nuclear spin that differ by $|\Delta m_I| = 2$ become mixed and the entire ensemble's dynamics at the corresponding crossings is significantly enhanced (Fig.~\ref{fig:quad}a). 
The odd transitions should still not be allowed. 
One has to consider that there is 
a small misalignment between the uniaxial symmetry of the ligand field and the quadrupolar interaction which leads to terms of the form: $I_zI_{\pm}$.
Figure~\ref{fig:quad} shows the effect of the above described non-axial contributions. It can be seen that they act mostly independently of each other, as the biaxial term mixes the hyperfine states with $|\Delta m_I| = 2$ and the angle deviation mixes the states with $|\Delta m_I| = 1$.

Crystal defects, solvent disorder, lattice mismatch between the YPc$_2$ and TbPc$_2$ molecules, and also the presence of a radical electronic spin non-uniformly distributed on the phthalocyanine ligands, are all factors that can lead to significant inhomogeneities in the electric field that can couple to the quadrupolar moment of the $^{159}$Tb nucleus.
However, in order to single out a dominant factor(s) that can result in a non-collinear quadrupolar interaction, further investigations are required.

\section{Conclusion}
In this article 
we report advancements in understanding
the resonant tunneling dynamics in the TbPc$_2$ complex as we take
an in depth look at interactions that can promote 
transitions between the hyperfine states. 
The rare combination of 
strong uniaxial character of the ligand field with
the tight hyperfine coupling results in characteristics that are
substantially different from the ones in transition metal ion molecular compounds~\cite{liu2014microscopic}.
We find that the nuclear spin dominates the dynamics of the molecular spin, as the usual suspects in the form of the environmental transverse field and ligand field fail to explain the transitions that do not conserve the nuclear spin. 
To explain why we see all the transitions (except at $H_z=0$~T) in the low temperature hysteresis loop we propose in in the non-axial quadrupolar interaction a mechanism that mixes the hyperfine states.
This result is important both from an academic point of view as similar dynamics can be observed in other lanthanide single molecule magnets, and from a technological one as tunneling between hyperfine states can be used to initialize and read-out the nuclear spins when implementing quantum protocols.

\section{Acknowledgement}
We acknowledge the Alexander von Humboldt Foundation and the ERC advanced grant MoQuOS No. 741276.
\bibliographystyle{apsrev4-1}
\bibliography{bib}

\end{document}